\def\scinot#1.{\hbox{$\,$ $\times$ $10^{#1}$}}

\documentclass{rsproca_fix}

\def\spose#1{\hbox to 0pt{#1\hss}}

\def\deg{\ifmmode^\circ\else$\null^\circ$\fi}
\def\gta{\mathrel{\spose{\lower 3pt\hbox{$\mathchar "218$}}\raise 2.0pt\hbox{$\mathchar"13E$}}}
\def\lta{\mathrel{\spose{\lower 3pt\hbox{$\mathchar "218$}}\raise 2.0pt\hbox{$\mathchar"13C$}}}

\begin{document}

\title{Atmospheric chemistry on Uranus and Neptune}

\author{
J. I. Moses$^{1}$, T. Cavali{\'e}$^{2,3}$, L. N. Fletcher$^{4}$, and M. T. Roman$^{4}$}

\address{$^{1}$Space Science Institute, 4765 Walnut St., Suite B, Boulder, CO 80301, USA\\
$^{2}$Laboratoire d'Astrophysique de Bordeaux, University of Bordeaux, CNRS, B18N, all{\'e}e Geoffroy Saint-Hilaire, 33615 Pessac, France\\
$^{3}$LESIA, Observatoire de Paris, 92195 Meudon, France \\
$^{4}$University of Leicester, School of Physics and Astronomy, University Road, Leicester, LE1 7RH, UK}

\subject{Uranus; Neptune; Ice Giants; Planetary Atmospheres; Atmospheric Chemistry}

\keywords{Uranus; Neptune; Photochemistry; Atmospheric Chemistry; Planetary Atmospheres}

\corres{Julianne I. Moses\\
\email{jmoses@spacescience.org}}


\begin{abstract}
Comparatively little is known about atmospheric chemistry on Uranus and Neptune, because remote spectral observations of these cold, 
distant ``Ice Giants'' are challenging, and each planet has only been visited by a single spacecraft during brief flybys in the 1980s.  
Thermochemical equilibrium is expected to control the composition in the deeper, hotter regions of the atmosphere on both planets, 
but disequilibrium chemical processes such as transport-induced quenching and photochemistry alter the composition in the 
upper atmospheric regions that can be probed remotely.  
Surprising disparities in the abundance of disequilibrium chemical products between the two planets point to significant differences in 
atmospheric transport.  
The atmospheric composition of Uranus and Neptune can provide critical clues for unravelling details of planet formation and evolution, but 
only if it is fully understood how and why atmospheric constituents vary in a three-dimensional sense and how material coming in from outside 
the planet affects observed abundances.  Future mission planning should take into account the key outstanding questions that remain 
unanswered about atmospheric chemistry on Uranus and Neptune, particularly those questions that pertain to planet formation and evolution, and 
those that address the complex, coupled atmospheric processes that operate on Ice Giants within our solar system and beyond.
\end{abstract}


\begin{fmtext}
\end{fmtext}

\maketitle

\section{Introduction}

Uranus and Neptune have the dubious honor of being the least explored planets in our solar system.  
Even after the \textit{Voyager 2} encounter
with Uranus in 1986 and Neptune in 1989, many mysteries remain regarding the 
interior structure, atmospheric properties, and magnetic-field generation on these planets.
It is also unclear how these Ice Giants formed.  
Traditional core-accretion planetary-formation models have difficulty getting the timing of 
Uranus and Neptune formation to work out correctly \cite{pollack96,helled14uran,helled14ppvi}.  
The main problem with these traditional models is that the planets tend to either grow too slowly to accrete much of the nebular 
gas in the protoplanetary disk before the gas is lost from the system, and so Uranus and Neptune end up as small solid planets 
without H$_2$-He envelopes, or the planets grow so rapidly that they reach a runaway gas-accretion stage, and so they
end up as H$_2$-He dominated gas giants.
Forming our Ice Giants with a substantial fraction of both heavy elements and H$_2$-He requires precise timing and 
fine-tuning of the models.  
This difficulty has become exacerbated in the current exoplanet-discovery era, given that
planets between the size of Earth and Neptune constitute a dominant fraction of the extrasolar planet population 
discovered to date \cite{batalha13,fressin13,petigura13}.  
We need to better understand how these volatile-rich, intermediate-sized planets form.
The observable atmospheric composition of Uranus and Neptune --- both from 
remote observations and future \textit{in situ} probes --- can provide important clues to formation location, formation mechanisms, protoplanetary 
disk conditions, and potential planetary migration history \cite{mousis18,hofstadter19,mandt20}, thereby considerably advancing our 
knowledge of planet formation and evolution.  However, the atmospheric composition of the Ice Giants can only provide these critical clues if 
we fully understand how various chemical processes are affecting the observed atmospheric composition, and how that observed composition 
relates to the bulk atmospheric composition.  

Observations of atmospheric constituents on Uranus and Neptune also shed light on the complex physical and chemical processes currently 
operating within the atmospheres, thus addressing the important broad topic of ``How planets work.''  Chemistry controls the 
atmospheric composition, which in turn affects other aspects of the atmosphere, such as cloud and haze formation, thermal structure and radiative 
balance, and atmospheric dynamics and circulation.  The three-dimensional distribution of atmospheric constituents --- and its variation with 
time --- reflect a complex coupling between atmospheric chemistry, dynamics, and energy transport in planetary atmospheres.  Observations and 
models can help work out the details of this coupling.  Some constituents may be good tracers for atmospheric dynamics, for example, but 
only if the sources, sinks, and overall chemistry of these species are well understood.  The two broadly similar Ice Giants also provide a
good launching point for comparative planetology.  By studying the composition of both Uranus and Neptune, we can identify the physical and 
chemical characteristics that are responsible for their observed similarities and differences.  The insight gained from this 
comparative-planetology approach may have relevance to studies of extrasolar planets.

In this paper, we review what is known and unknown about atmospheric chemistry on Uranus and Neptune.  That chemistry has many similarities 
to Jupiter and Saturn.  However, some differences arise among all our giant planets as a result of heavy-element content, atmospheric 
temperatures, atmospheric dynamics and circulation, and the stochastic nature of some processes, such as impacts, which can affect planetary obliquity, 
interior structure, interior mixing, internal heat flow, deep atmospheric convection, and atmospheric composition 
\cite{safronov66,stevenson86,podolak12,kegerreis18,reinhart20}.  
We will discuss the observed composition of Uranus and Neptune, as well as the key chemical processes that operate in various regions of 
the atmosphere, from the deep troposphere on up to the top of the atmosphere.
Our \textit{Voyager}-era understanding of atmospheric chemistry on Uranus and Neptune has been described in previous works
\cite{strobel91,atreya91,fegley91,bishop95,gautier95}; here, we restrict our discussion to the major advances that have occurred 
since the mid-1990s.

\section{Basic atmospheric properties and observed composition}

\begin{figure}[!ht]
\centering\includegraphics[width=5.2in]{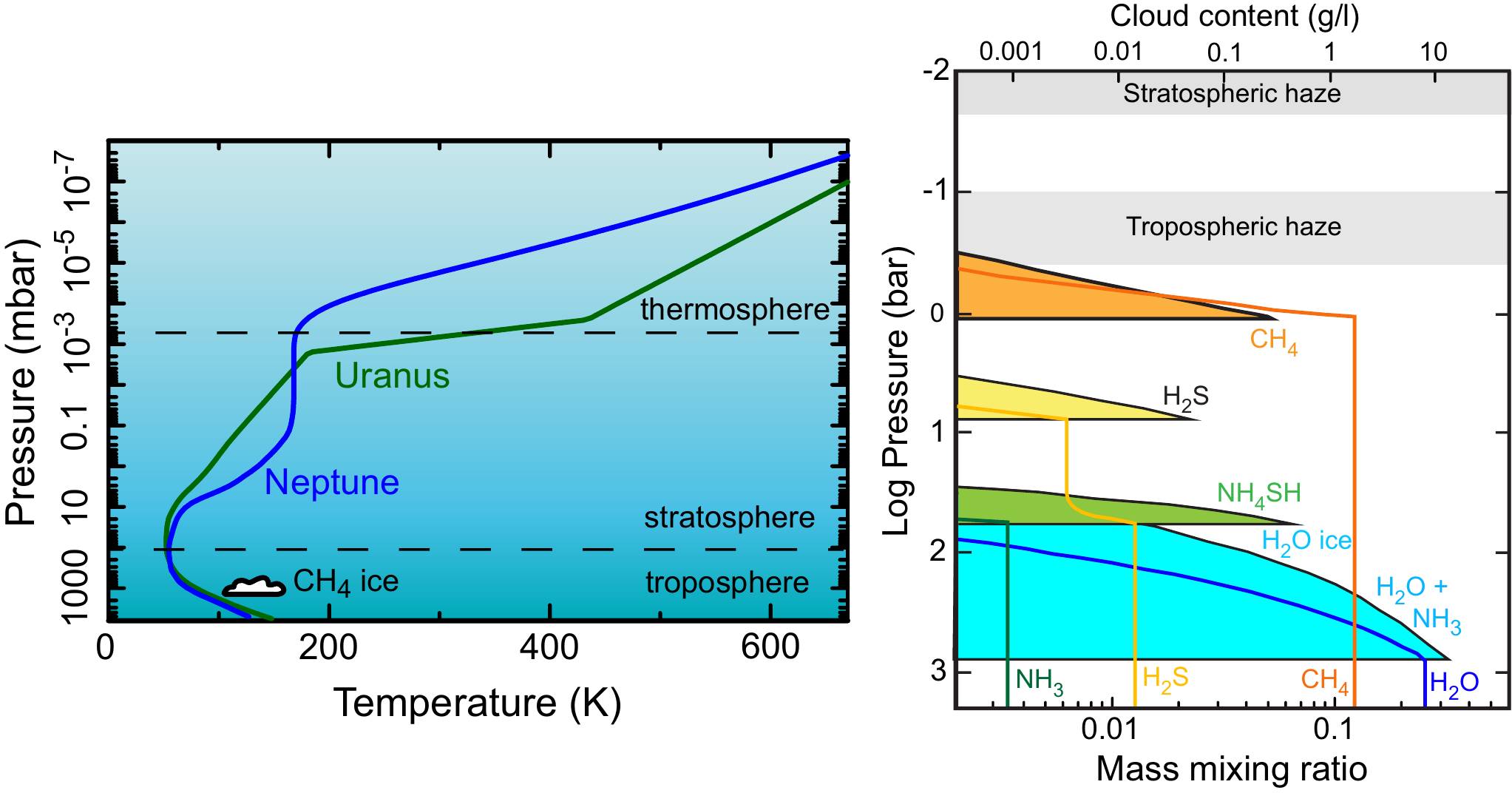}
\caption{(Left) Global-average temperature-pressure profile of the atmospheres of Uranus and Neptune, with major regions of the 
atmosphere labeled (modified from Moses et al. \cite{moses18}).  (Right) Thermochemical equilibrium prediction of the upper-tropospheric cloud structure 
on Uranus (modified from Hueso and S{\'a}nchez-Lavega \cite{hueso19}).  The predicted 
mass mixing ratios of condensible gases are shown as colored solid lines, and the maximum cloud density as solid black lines with color-shaded regions.}
\label{figtempclouds}
\end{figure}

The bulk composition of Uranus and Neptune by mass is roughly 10-20\% hydrogen and helium and 80-90\% heavier elements \cite{podolak19}.  
These heavier elements include refractory elements that traditionally form rocky materials and more volatile elements that traditionally 
form ices, but the relative proportions of ``rock'' and ``ice'' are unclear for both the planets as a whole, as well as for their 
atmospheres.  
What is clear is that the atmospheres themselves becomes progressively enriched with H$_2$ and He with increasing altitude as elements condense out 
and are thereby removed from cooler higher-altitude regions, and then more significantly enriched in H$_2$ and He at very high altitudes as mass 
separation occurs above the homopause.  The atmospheres of Uranus and Neptune can be divided into 
three main regions: the \textit{troposphere} in the lowest region, where convection occurs and temperatures are hot at depth and decrease 
with increasing altitude, the \textit{stratosphere} or middle atmosphere above the tropopause temperature minimum, where radiative 
processes dominate and temperatures increase or are constant with increasing altitude, and the \textit{thermosphere} in the uppermost 
regions of the atmosphere, where temperatures begin to increase more sharply as heat is conducted downward from a hot exosphere at high 
altitudes (see Fig.~\ref{figtempclouds}).

Observations at various wavelengths can probe down to the few-bar level in the upper troposphere, but gas and cloud opacity and Rayleigh scattering 
limit the penetration any deeper \cite{hueso19}.  The expected upper-tropospheric cloud structure from thermochemical-equilibrium arguments is 
shown in Fig.~\ref{figtempclouds}, as is described in Hueso et al.~\cite{hueso19}.  This figure illustrates the cloud 
structure for Uranus, but the results for Neptune are similar.  In this thermochemical-equilibrium scenario, a liquid 
water and ammonia solution cloud forms at the several-hundred bar level, transitioning to a water-ice cloud near its top.  Gas-phase H$_2$O 
is strongly depleted in this region as the water condenses.  Above the water cloud, ammonia (NH$_3$) and hydrogen sulfide (H$_2$S) can combine to 
form an ammonium hydrosulfide (NH$_4$SH) cloud, substantially depleting whichever gas phase species 
is less abundant.  Observations of both Uranus and Neptune suggest that H$_2$S is more abundant than NH$_3$ 
\cite{gulkis78,briggs89,depater91,irwin18,irwin19neph2s,tollefson19}.  The remaining H$_2$S then forms an H$_2$S ice cloud at the several-bar level, 
and methane (CH$_4$) then condenses as an ice cloud near the $\sim$1-bar level.  Various optically thin hazes from disequilibrium species and 
photochemical products reside above the CH$_4$ cloud.  Below the water cloud, other equilibrium cloud layers are predicted to form \cite{fegley85uran,fegley86}, 
depleting the atmosphere in refractory elements above their condensation regions.  In fact, of all the equilibrium species predicted to be present in the 
atmospheres of Uranus and Neptune, only H$_2$, He, CH$_4$, and H$_2$S are directly detected spectroscopically, while everything else that has been observed 
is believed to be produced from disequilibrium chemical processes.

\begin{table}[!htb]   
\caption{Tropospheric composition by volume (above the water-solution cloud) from selected recent references}
\label{tabtrop}
\begin{tabular}{llll}
\hline
species & Uranus & Neptune & notes/references \\
\hline
He     & 15.2\%        & 14.9\%      & \cite{conrath87,burgdorf03} \\
CH$_4$ & 1.4--4\%      & 2--5\%      & latitude dependent; \cite{karkoschka09,karkoschka11,irwin19nepch4,sromovsky14} \\
NH$_3$ & 30--90 ppm    & 40--200 ppm & inferred from microwave photometry; \cite{hofstadter18,depater91,deboer96} \\
H$_2$O & $<$ 5\% $^\star$ & 27\% $^\star$  & $^{\star}$indirect determination of deep abundance; \cite{venot20fixed,cavalie17} \\
PH$_3$ & $<$ 2 ppm     & $<$ 1.1 ppb at 0.7 bar & \cite{encrenaz96,teanby19} \\
H$_2$S & 0.4--0.8 ppm  & 1--3 ppm    & at H$_2$S cloud top; \cite{irwin18,irwin19neph2s} \\
       & $>$ 10-25 ppm & 700 ppm     & below H$_2$S cloud (latitude dependent); \cite{irwin18,tollefson19} \\
\hline
\end{tabular}
\vspace*{-4pt}
\end{table}

\begin{table}[!htb]
\caption{Stratospheric composition by volume from selected recent references}
\label{tabstrat}
\begin{tabular}{llll}
\hline
species & Uranus & Neptune & notes/references \\
\hline
CH$_4$     & 16 ppm at 50 mbar    & 0.115\% at 5 mbar & \cite{orton14temp,lellouch15} \\
C$_2$H$_2$ & 0.25 ppm at 0.2 mbar & 0.033 ppm at 0.5 mbar & \cite{orton14chem,greathouse11} \\
C$_2$H$_4$ & $<$ 2$\scinot-14.$ at 10 mbar & 0.8 ppb at 0.2 mbar & \cite{orton14chem,schulz99} \\
C$_2$H$_6$ & 0.13 ppm at 0.2 mbar & 0.85 ppm at 0.3 mbar & \cite{orton14chem,fletcher10akari} \\
C$_3$H$_4$ & 0.36 ppb at 0.4 mbar & 0.12 ppb at 0.1 mbar & \cite{orton14chem,meadows08} \\
C$_4$H$_2$ & 0.13 ppb at 0.4 mbar & 0.003 ppb at 0.1 mbar & \cite{orton14chem,meadows08} \\
CO$_2$     & 0.08 ppb at 0.14 mbar & 0.78 ppm at 0.1 mbar & external source; \cite{orton14chem,meadows08} \\
CO         & 6 ppb at 0.5 mbar    & 1.1 ppm at 0.1 mbar & external source; \cite{cavalie14,luszczcook13} \\
H$_2$O     & 3.8 ppb at 0.03 mbar    & 2.5 ppm at 0.16 mbar & external source; \cite{feuchtgruber99} \\
D/H        & 4.4$\scinot-5.$      & 4.1$\scinot-5.$     & from HD; \cite{feuchtgruber13} \\
\hline
\end{tabular}
\vspace*{-4pt}
\end{table}

The atmospheric constituents observed during the \textit{Voyager 2} era and earlier are discussed in several reviews 
\cite{strobel91,atreya91,fegley91,bishop95,gautier95}.  Since that time, H$_2$S has been definitively detected in the troposphere of Uranus 
and tentatively on Neptune \cite{irwin18,irwin19neph2s}, several new disequilibrium products and exogenic species have been detected throughout 
the atmosphere \cite{rosenqvist92,marten93,trafton93,feuchtgruber97,feuchtgruber99,bezard99,schulz99,encrenaz04,burgdorf06,meadows08,orton14chem,moreno17}, 
and refinements to previous abundance measurements have been reported 
\cite{guilloteau93,naylor94,courtin96,encrenaz96,lellouch05,marten05,hesman07,fletcher10akari,lellouch10,luszczcook13,irwin14,teanby19,bishop92,orton92,lellouch94,encrenaz98,bishop98,feuchtgruber99hd,trafton99,encrenaz00,encrenaz03,feuchtgruber03,burgdorf03,fouchet03,hammel06,cavalie08,karkoschka09,greathouse11,karkoschka11,melin11uran,irwin12new,feuchtgruber13,teanby13,fletcher14nep,orton14temp,rezac14,iino14,cavalie14,lellouch15,fletcher18,sromovsky19,tollefson19}.
Some spatially resolved information on species abundances has also become available in recent years
\cite{karkoschka09,greathouse11,karkoschka11,sromovsky11,irwin12new,luszczcook13spatres,tice13,fletcher14nep,sromovsky14,depater14,roman18,irwin18,sromovsky19,tollefson19,irwin19neph2s,irwin19nepch4,roman20}.
Table~\ref{tabtrop} and Table~\ref{tabstrat} summarize the observed abundance of tropospheric and stratospheric constituents on the Ice Giants.  Values in these 
tables have been chosen from a single reference observation per planet; for additional observations and more in-depth discussions of uncertainties, see the full 
suite of references listed above.


\section{Tropospheric chemistry}

Thermochemical equilibrium dominates the chemistry in the deep troposphere on Uranus and Neptune, and transport-induced quenching and photochemistry 
affect the upper troposphere.

\subsection{Thermochemical equilibrium and quenching}

Deep in the tropospheres of Uranus and Neptune at pressures greater than several kilobar, temperatures are high enough and reaction rates are fast 
enough that thermochemical equilibrium can be 
maintained.  The deep tropospheric composition can then be predicted as a function of pressure, temperature, and elemental abundances by assuming 
thermochemical equilibrium \cite{fegley85uran,fegley86}.  However, gas that is transported upward from the deep troposphere will cool, and the colder 
temperatures will inhibit some chemical reactions (e.g., those with high-energy barriers), such that thermochemical equilibrium will 
eventually cease to be maintained kinetically.  When the rate of vertical transport exceeds the rate at which chemical reactions convert 
between different molecular forms of an element, then the composition can be ``frozen in'' at that stage \cite{prinn77}, producing vertically 
uniform mixing ratios of the involved species above the quench point, in the absence of any other chemical processes such as 
photochemistry or condensation.  As a consequence of this disequilibrium 
transport-induced quenching process, species such as CO, N$_2$, PH$_3$, GeH$_4$, C$_2$H$_6$, HCN, HCl, HF, CO$_2$, CH$_3$OH, CH$_3$SH, CH$_3$NH$_2$, and H$_2$Se 
are expected to survive into the upper troposphere of Uranus and Neptune at abundances much greater than equilibrium predictions \cite{fegley86}.
Because the quenched abundances depend on the bulk elemental abundances in the deep atmosphere, as well as the strength of atmospheric mixing, 
these disequilibrium species are potentially important indicators of conditions in the deep tropospheres of the Ice Giants, where both \textit{in situ} 
probes and remote-sensing observations cannot penetrate.  Of the aforementioned disequilibrium species, only carbon monoxide (CO) has been definitively 
detected in the \textit{troposphere}, and only on Neptune \cite{rosenqvist92,marten93,lellouch05,hesman07,luszczcook13,teanby13,teanby19}. However, even upper 
limits for these species can provide important information about the deep atmosphere \cite{fegley91}.

\begin{figure}[!ht]
\centering\includegraphics[width=5.2in]{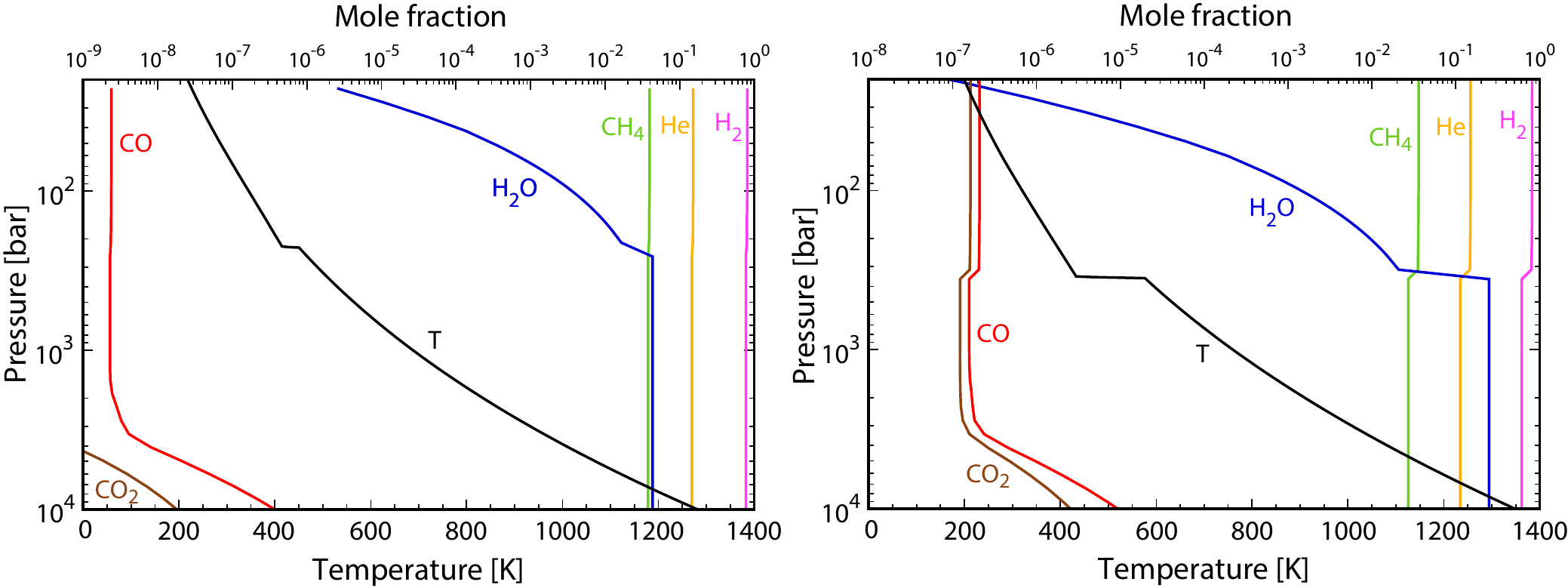}
\caption{Mole fraction profiles of some key species in the deep troposphere of Uranus (Left) and Neptune (Right), as predicted from a thermochemical 
kinetics and diffusion model \cite{cavalie17,venot20fixed}.  The deep carbon and oxygen abundances are varied until the model reproduces the observed upper 
tropospheric CH$_4$ mixing ratio at the equator \cite{karkoschka09,karkoschka11} and the observed upper-tropospheric CO mixing ratio (Neptune) or CO 
upper limit (Uranus) \cite{teanby13,luszczcook13}. Figure modified from Venot et al. \cite{venot20fixed}.}
\label{figcavdeep}
\end{figure}

Carbon monoxide has also been detected in the stratospheres of Uranus and Neptune with a mixing ratio greater than is found in the troposphere 
\cite{lellouch05,hesman07,teanby13,cavalie14}, 
indicating that CO has an external source on these planets.  However, tropospheric CO can also result from an internal, deep quenched source. 
The observed upper-tropospheric CO volume mixing ratio on Neptune is surprisingly large, but somewhat uncertain --- various reports put it in the 
0.08--1.5 ppm range
\cite{rosenqvist92,marten93,guilloteau93,naylor94,courtin96,encrenaz96,lellouch05,marten05,hesman07,fletcher10akari,lellouch10,luszczcook13,irwin14,teanby19}.
The uncertainty arises from observational difficulties and model dependencies in the fitting of the pressure-broadened wings 
of the CO absorption features.  In contrast, the most stringent upper limit for tropospheric CO on Uranus is as low as 2.1$\scinot-9.$ \cite{teanby13} --- 
a low value that presumably results from Uranus' expected less efficient tropospheric mixing (as supported by its low internal heat flux [103]) and 
potentially smaller intrinsic deep oxygen abundance \cite{cavalie17,venot20fixed}.
Observational constraints on the tropospheric CO abundances on Uranus and Neptune have been used to constrain the deep O/H 
abundance on these planets, assuming that the CO derives from quenching from the deep interior \cite{fegley91,lodders94,luszczcook13,cavalie14,cavalie17,venot20fixed}.  
Early estimates via time-scale arguments \cite{fegley91,marten93,lodders94,luszczcook13} have given way to more complete thermochemical kinetics and 
transport models \cite{cavalie14,cavalie17,venot20fixed}, based on those developed for Jupiter and extrasolar planets \cite{visscher10co,moses11,visscher11,venot12}. 
The full suite of these estimates suggests a deep oxygen enrichment relative to hydrogen on Neptune of 250--650 times solar, while the upper limit for 
Uranus is $<$ 45--260 times solar.  

Figure~\ref{figcavdeep} shows the predicted vertical profiles of a few species in the deep troposphere of Uranus and Neptune from the nominal thermochemical 
kinetics and diffusion model of Venot et al. \cite{venot20fixed}, updated from Cavali{\'e} et al. \cite{cavalie17}.  Both planets are assumed to have a 
deep-tropospheric eddy diffusion coefficient $K_{zz}$ = 10$^8$ cm$^{2}$ s$^{-1}$ in this model, as estimated from mixing-length theory (for which $K_{zz}$ 
varies with the internal heat flux to the one-third power, so the factor of $\sim$10 difference in the internal heat flux of the two planets translates to only 
a factor of $\sim$2 difference in $K_{zz}$, although it is possible that convection is inhibited on Uranus \cite{vazan20}, making $K_{zz}$ even smaller).   
Venot et al.~find that a deep O/H abundance of 250 times solar for Neptune and $<$ 45 times solar for Uranus 
are needed to reproduce the upper-tropospheric observational constraints 
for CO \cite{teanby13,luszczcook13}.  As discussed below, this solution is somewhat model dependent.  
Other interesting features of the deep tropospheres of the Ice Giants are apparent from Fig.~\ref{figcavdeep}.  The large discontinuity that appears 
in the temperature profile and species abundances in the middle troposphere is a result of water condensation.  Water is a major component of the Ice Giants, and 
when H$_2$O is removed from the gas phase due to condensation, the mixing ratios of all the remaining gas-phase species --- including H$_2$ --- increase 
notably.  The mean molecular mass of the atmosphere also drops significantly when H$_2$O condenses, which can produce a stable layer that inhibits 
convection \cite{leconte17,cavalie17}.  These types of models also predict 
the quench behavior of CO$_2$, CH$_3$OH, N$_2$, HCN, and other constituents \cite{moses10,moses11,venot12,cavalie14,cavalie17,venot20fixed}, which could 
in theory supply additional constraints on the deep elemental abundances and/or vertical transport rates if these species are ever detected in Ice-Giant 
tropospheres \cite{fegley91}.

Thermochemical kinetics and transport models therefore have the potential for being a powerful tool for indirectly determining the bulk elemental 
atmospheric abundances on the Ice Giants (see also the review of Cavali{\'e} et al.~\cite{cavalie20}).  Many of the assumptions and inputs to these 
models are uncertain, however, adding significant uncertainties to the final result.
For example, the model results are sensitive to the thermal structure of the deep atmosphere, which is not well constrained and must be extrapolated many orders 
of magnitude from sparse measurements from the upper troposphere, and which is affected by complicated and often poorly characterized physical and chemical effects 
(e.g., non-ideal gas effects, critical-point behavior, wet-versus-dry adiabats, double-diffusive convection, the atmospheric evolution history, 3D dynamical 
effects) \cite{lodders94,leconte17,cavalie17}.  Vertical transport is typically parameterized through an eddy diffusion coefficient profile 
in these models, which can depend on the internal heat flux, temperature profile, and composition profile, and is predicted to be latitude dependent 
\cite{gierasch85,visscher10co,wang15deep}.  The model results are also very sensitive to individual chemical reaction rates involved in the conversion between 
different forms of an element, the uncertainties of which can lead to potential order-of-magnitude uncertainties in the predicted abundance of quenched 
species \cite{venot12,moses14,wang16}.  When Venot et al. \cite{venot20fixed} updated the reaction mechanism used in Cavali{\'e} et al. \cite{cavalie17}, for 
instance, their revisions led to a reduction in their previously predicted values of the deep oxygen abundance by a factor of 2 for Neptune and a factor
of 3.5 in the upper limit for Uranus.
Judging from comparisons with other combustion-based and atmospheric-based kinetics models 
\cite{moses14,wang16}, the uncertainties in the deep O/H abundance that result from kinetics uncertainties could be even larger than this factor of 
a few.  However, if the above model uncertainties could be reduced, this indirect means of determining deep elemental abundances on the Ice Giants holds 
much promise, particularly if future observations and missions are able to measure the tropospheric abundance of multiple disequilibrium quenched species 
\cite{visscher05,cavalie20}.

\subsection{Tropospheric photochemistry}

Longer-wavelength ultraviolet photons from the Sun can reach the upper troposphere to initiate local photochemistry, although Rayleigh scattering and aerosol 
extinction ultimately limit how far the photons can penetrate.  Tropospheric photochemistry on Uranus and Neptune has not received much attention, in part 
because of a lack of observational motivators.  Although phosphine (PH$_3$) --- believed to be a disequilibrium quenched species that should be the dominant 
form of phosphorous in the upper troposphere \cite{fegley85uran} --- can condense in the cold upper tropospheres of Uranus and Neptune, PH$_3$  may be 
photolyzed and destroyed before condensation can occur.  The resulting photochemistry would lead primarily to the formation of P$_2$H$_4$ hazes, as has been 
suggested for Saturn \cite{kaye84ph3}.  Indeed, PH$_3$ photolysis is efficient in the recent Neptune tropospheric photochemical model presented by Teanby et al. 
\cite{teanby19} and can easily explain their low derived 1.1 ppb upper limit for PH$_3$ near 0.4--0.8 bar.  Why, then, does photolysis not remove PH$_3$ 
from view on Jupiter and Saturn?  The difference in upper tropospheric PH$_3$ abundance on these planets is the result of enhanced shielding of the PH$_3$ 
from photolysis on Jupiter and Saturn due to the presence of NH$_3$ gas and cloud particles in the upper troposphere, whereas NH$_3$ has already been removed 
at deeper pressures on Neptune (see Fig.~\ref{figtempclouds}).  The rate of vertical mixing in the upper tropospheres of these planets may also play a role.
The derived PH$_3$ profile from the Teanby et al.~model suggests 
that any atmospheric probe would need to go to at least 10 bar to have a chance of sampling the deep phosphorus abundance, due to the vertical gradient imposed by 
strong photochemical loss, slow mixing, and potential condensation in the radiative portion of the upper troposphere.  The expected low NH$_3$ abundance 
in the 1--3 bar pressure region where PH$_3$ is being photolyzed (see Fig.~\ref{figtempclouds}) limits the importance of the coupled NH$_3$--PH$_3$ photochemistry 
that is suggested to occur on Jupiter and Saturn \cite{moses04,fouchet09}.  Similarly, the expected low abundance of hydrocarbon radicals and unsaturated hydrocarbons 
in the region where PH$_3$ is being photolyzed limits the coupling of PH$_3$ and hydrocarbon photochemistry.  Under favorable conditions of strong convective 
uplift, H$_2$S may be carried to high-enough altitudes to interact with photons of less than $\sim$260 nm, which can photolyze the H$_2$S.  Under these 
rare but interesting conditions, sulfur photochemistry might occur.  The details of H$_2$S photochemistry under reducing conditions are poorly understood, 
due to a lack of relevant chemical kinetics data, but condensed elemental sulfur is a potential end product of this chemistry.
  
\subsection{Tropospheric latitude variations}

\begin{figure}[!ht]
\centering\includegraphics[width=5.2in]{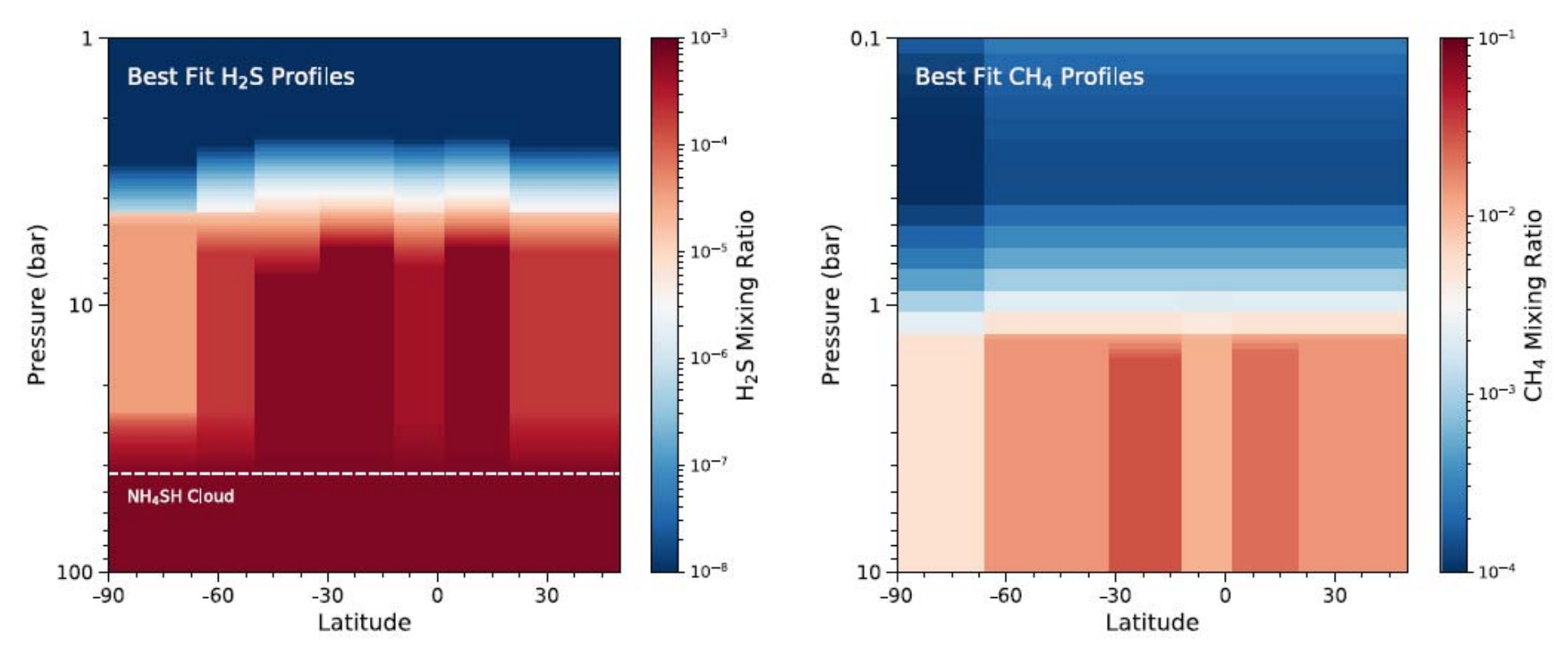}
\caption{Mixing ratios of H$_2$S (Left) and CH$_4$ (Right) as a function of latitude on Neptune from models that provide the best fits to the ALMA data 
of Tollefson et al. \cite{tollefson19}.  Figure from \cite{tollefson19}.}
\label{figtollefson}
\end{figure}

In recent years, observations have demonstrated that tropospheric condensible species such as CH$_4$ and H$_2$S have abundances that vary with latitude 
on both Uranus and Neptune 
\cite{karkoschka09,karkoschka11,sromovsky11,irwin12new,tice13,sromovsky14,roman18,irwin18,sromovsky19,tollefson19,irwin19neph2s,irwin19nepch4,roman20} (see
Fig.~\ref{figtollefson}).
These species have strong, broad-scale, equator-to-pole mixing-ratio gradients, with higher abundances at low latitudes than high latitudes, and some 
evidence for meridional variability on smaller scales.  In contrast, 
spatially resolved observations of Jupiter and Saturn show variations in tropospheric NH$_3$ that occur on more local scales due to the belt-zone structure 
\cite{fletcher11vims,janssen13,bolton17sci,li17juno,giles17,blain18,depater19}.  Latitude variations in tropospheric condensible species could be common on all the 
giant planets, with the \textit{Juno} observations in particular demonstrating that abundances can be variable  
to much deeper pressures than previously realized \cite{bolton17sci,li17juno,li20}.  
The complete vertical profiles of CH$_4$ and H$_2$S on Uranus and Neptune as a function of latitude have not yet been worked out, nor 
have NH$_3$ and H$_2$O been definitively detected.
An important goal for future observations and missions is the determination of how the temperatures and the gas-phase abundances of these condensible 
species vary in a three-dimensional sense.  The distributions of condensible species are intimately linked to atmospheric 
dynamics.  The observed variation 
is useful for determining broad-scale atmospheric circulation on the Ice Giants \cite{depater14,fletcher20beltzone,fletcher20circ}, as well as for 
furthering our understanding of how convection originates and operates in hydrogen-dominated atmospheres for which the condensate is heavier than the 
background gas \cite{guillot19}.  Moist air in H$_2$-dominated atmospheres will tend to sink rather than rise, but latent-heat release from condensation 
\cite{stoker86,lunine89} and energy release from ortho-para H$_2$ conversion \cite{smith95} can fuel storms.  How 
moist convection is initiated in the methane cloud layer on the Ice Giants is not immediately obvious \cite{stoker89,lunine89,hueso19}, nor is it clear 
how important localized storms are compared to broader circulation in controlling condensible species abundances, atmospheric structure, and energy transport on 
hydrogen-dominated planets \cite{gierasch00,sugiyama14,li17juno,ingersoll17,fletcher20circ,fletcher20beltzone,depater19,guillot19}.  

Latitude and other spatial variations complicate analyses of remote-sensing data and plans for future deep-probe missions.  The 
degeneracy between temperatures and species abundances in atmospheric retrievals will be even more difficult to break when both are non-uniform 
across the planet.  Given that species abundances vary spatially, where do we want to send potential probe(s)?  How deep do the
probes need to go to sample the well-mixed deep abundance of different species?

Tropospheric species that neither condense nor participate in active chemistry (e.g., noble gases and chemically long-lived species) 
are expected to have mixing-ratio profiles that are constant with 
altitude and latitude.  This theoretical expectation is not backed up by any current observations, but with no known production or loss processes, 
there is no apparent way to alter the volume mixing ratio in a background gas that remains stable.  Exceptions to this prediction are quenched 
disequilibrium species such as CO and N$_2$, whose mixing ratios could potentially vary with latitude as a result of meridional variations 
in the convective transport rates in the deep troposphere that affect the quench location \cite{wang16}.  Tropospheric constituents that might not 
vary with latitude on Uranus and Neptune, such as noble gases and their isotopes, are prime targets for measurements on future probe missions 
\cite{mousis19}.  The relative abundance of these gases can help identify the origin of the heavy elements in the Ice Giant 
atmospheres from various possible reservoirs in the protosolar nebula \cite{mousis18,mousis19,mandt20}.  Molecular nitrogen (N$_2$) and CO are important 
quenched species from the deep atmosphere that would help constrain the deep N/H and O/H abundances.  Given that CO has the same molecular weight as 
N$_2$, and the two would be indistinguishable from each other in a low-resolution mass spectrometer, a probe that has instrumentation that could differentiate 
between N$_2$ and CO would be ideal.

\section{Stratospheric chemistry}

Despite the weak solar ultraviolet flux in the outer solar system, photochemistry on Uranus and Neptune is quite active and vigorous.  
Methane photolysis initiates the stratospheric photochemistry on these planets, producing a slew of hydrocarbon photochemical products 
\cite{atreya91,bishop95}, many of which are observed (see Table~\ref{tabstrat}).  Seasonal changes in solar forcing are expected to lead to temporal and 
spatial differences in stratospheric chemistry. Surprising discrepancies in the abundance of photochemical products between the two planets point to 
significant differences in the strength of atmospheric mixing \cite{yelle87,herbert87,broadfoot89,moses18}.  External material coming into the atmosphere 
from interplanetary dust particles, comets, and the local satellite/ring system contributes to stratospheric chemistry 
\cite{feuchtgruber99}.  

\subsection{Methane photochemistry}

Although methane condenses in the upper troposphere of both planets, sufficient CH$_4$ is transported up into the stratosphere --- by moist convection 
or other processes that are not well understood --- where the methane can interact with solar photons of wavelengths less than $\sim$145 nm, 
triggering photolysis.  Photochemical models for the stratospheres of Uranus and Neptune have been presented by several groups 
\cite{atreya83,herbert87,romani88,romani89,summers89,bishop90,bishop92,moses92nucl,romani93,lellouch94,moses95c,dobrijevic98,bishop98,bezard99,schulz99,moses05,dobrijevic10a,orton14chem,cavalie14,moses17poppe,moses18,lara19,dobrijevic20}.
These models indicate that a variety of hydrocarbons are produced from stratospheric methane photochemistry, including methyl radicals (CH$_3$), 
acetylene (C$_2$H$_2$), ethylene (C$_2$H$_4$), ethane (C$_2$H$_6$), methylacetylene (CH$_3$C$_2$H), and diacetylene (C$_4$H$_2$), which have been observed 
on one or both of these planets (see Table~\ref{tabstrat}).  These products and others can be photolyzed themselves, leading to a complex and intricate 
hydrocarbon kinetics that has many similarities to that on Jupiter and Saturn (see the reviews of \cite{moses04,fouchet09}).  
The photochemical products will flow down from the methane photolysis region in the upper stratosphere to the lower stratosphere, 
where many of the products can condense to form hazes, as are observed in the $\sim$1--100 mbar region \cite{pollack87,pryor92,moses95c}.  
Ethane, acetylene, and diacetylene are the dominant photochemical components of this haze, 
but other hydrocarbons and externally supplied species such as H$_2$O, CO$_2$, and HCN --- the latter which could also derive from cosmic-ray dissociation 
of internal N$_2$ or from N from Triton \cite{marten93,lellouch94} --- also contribute to the stratospheric aerosol.  

\begin{figure}[!ht]
\centering\includegraphics[width=5.2in]{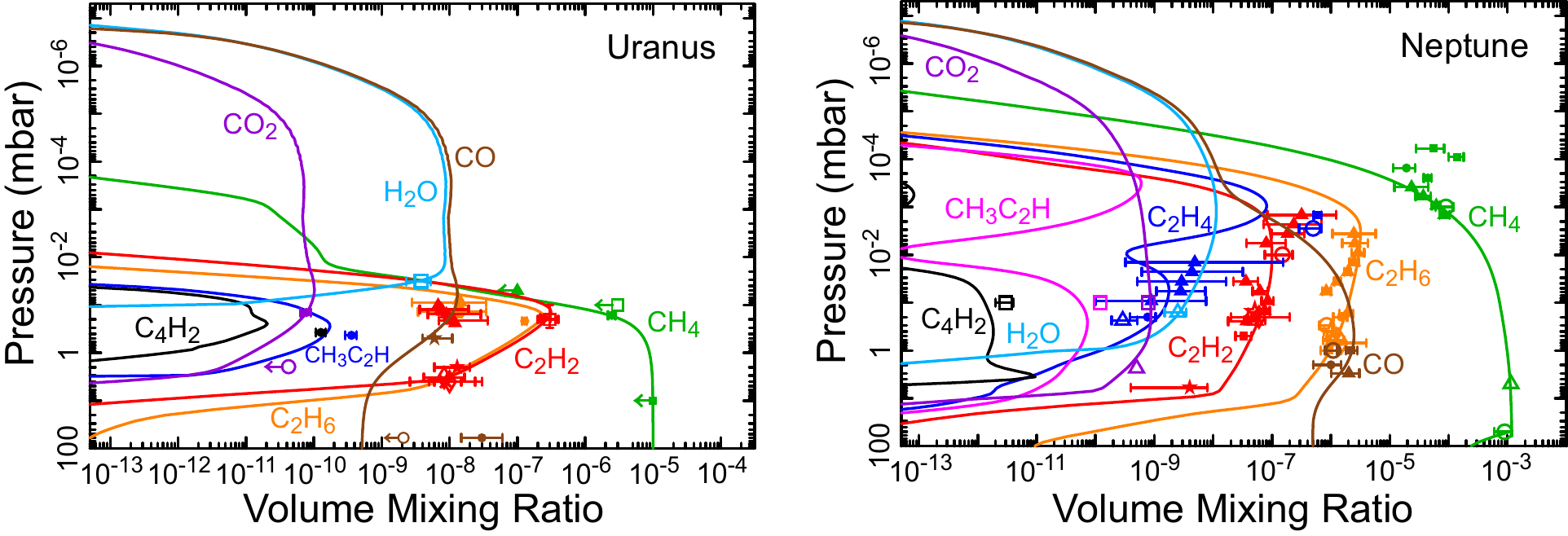}
\caption{Vertical mixing-ratio profiles for several stratospheric constituents on Uranus (Left) and Neptune (Right) as predicted from 1-D global-average 
photochemical models (colored lines), compared to various observations (data points with associated error bars).  The models for both planets include 
an external source of oxygen from the ablation of interplanetary dust \cite{moses17poppe}, while the Neptune model includes a source of CO from a large 
cometary impact 200 years ago.  The sharp drops in species mixing ratios in the lower stratosphere are caused by condensation.  Figure modified from 
Moses and Poppe \cite{moses17poppe}.}
\label{figchemurannep}
\end{figure}

Key differences between the photochemical product abundances on Uranus and Neptune arise predominantly from differences in the strength of 
stratospheric mixing between the two planets (see Fig.~\ref{figchemurannep}), which results from the presence and strength of atmospheric 
waves (including breaking gravity waves), circulation and winds, and turbulence of all scales.  Mixing is weak on Uranus but strong on Neptune.  
The stronger mixing on Neptune allows CH$_4$ to be carried to much lower pressures (higher altitudes) than on Uranus (see Fig.~\ref{figchemurannep}), 
before the homopause level is attained and molecular diffusion dominates, causing species such as methane and its photochemical products that 
are heavier than the background H$_2$ to drop off sharply with altitude. 
The higher homopause altitude on Neptune allows the photochemical products to build up over a much larger vertical column above their condensation 
regions in the lower stratosphere than on Uranus. 
The column abundances of photochemically produced hydrocarbons therefore tend to be larger on 
Neptune, and are easier to observe than on Uranus --- Neptune observations are also aided by a larger temperature gradient and warmer stratosphere.  
The pressure at which CH$_4$ is photolyzed also affects the subsequent photochemistry \cite{moses05,orton14chem}, leading to differences in the 
relative abundances of the hydrocarbon products on Uranus versus Neptune.  

Detailed descriptions of neutral hydrocarbon photochemistry on Uranus and Neptune are provided in \cite{moses05,orton14chem,moses18,dobrijevic20}.  
Dobrijevic et al. \cite{dobrijevic10a} discuss how uncertainties in reaction rate coefficients propagate to uncertainties in hydrocarbon 
abundances in photochemical models.  Model uncertainties are as large or larger than observational uncertainties for many species.

\subsection{Dependence on latitude and season}

The abundance of hydrocarbon photochemical products on Uranus and Neptune depends on latitude and season.  Uranus' extreme axial tilt of 
97.8$^{\circ}$ and Neptune's more moderate axial tilt of 28.3$^{\circ}$ cause seasonal variations in the solar actinic flux that drives photochemistry 
on the Ice Giants.  This obliquity, combined with the long orbital periods of the planets, ensures strong and long-lasting differences in the 
production and loss rates of hydrocarbons over time.  Portions of Uranus and Neptune experience many 
Earth years of darkness during winter, with only a small amount of solar Lyman alpha radiation scattered from hydrogen in the local interplanetary medium 
providing any source of photolyzing radiation.  Averaged over of a full planetary year, the poles of Uranus receive a greater flux of ultraviolet 
radiation that the equator, while the opposite is true for Neptune.  This seasonally variable solar forcing results in meridional gradients in 
photochemical product abundances that change with time and altitude.  Two-dimensional or three-dimensional time-variable models are needed to 
track this variation (e.g.,  \cite{hue18} for Jupiter).  Moses et al. \cite{moses18} have presented such two-dimensional time-variable photochemical 
models for Uranus and Neptune, under the assumption of rapid zonal homogenization and no meridional transport.

\begin{figure}[!ht]
\centering\includegraphics[width=5.2in]{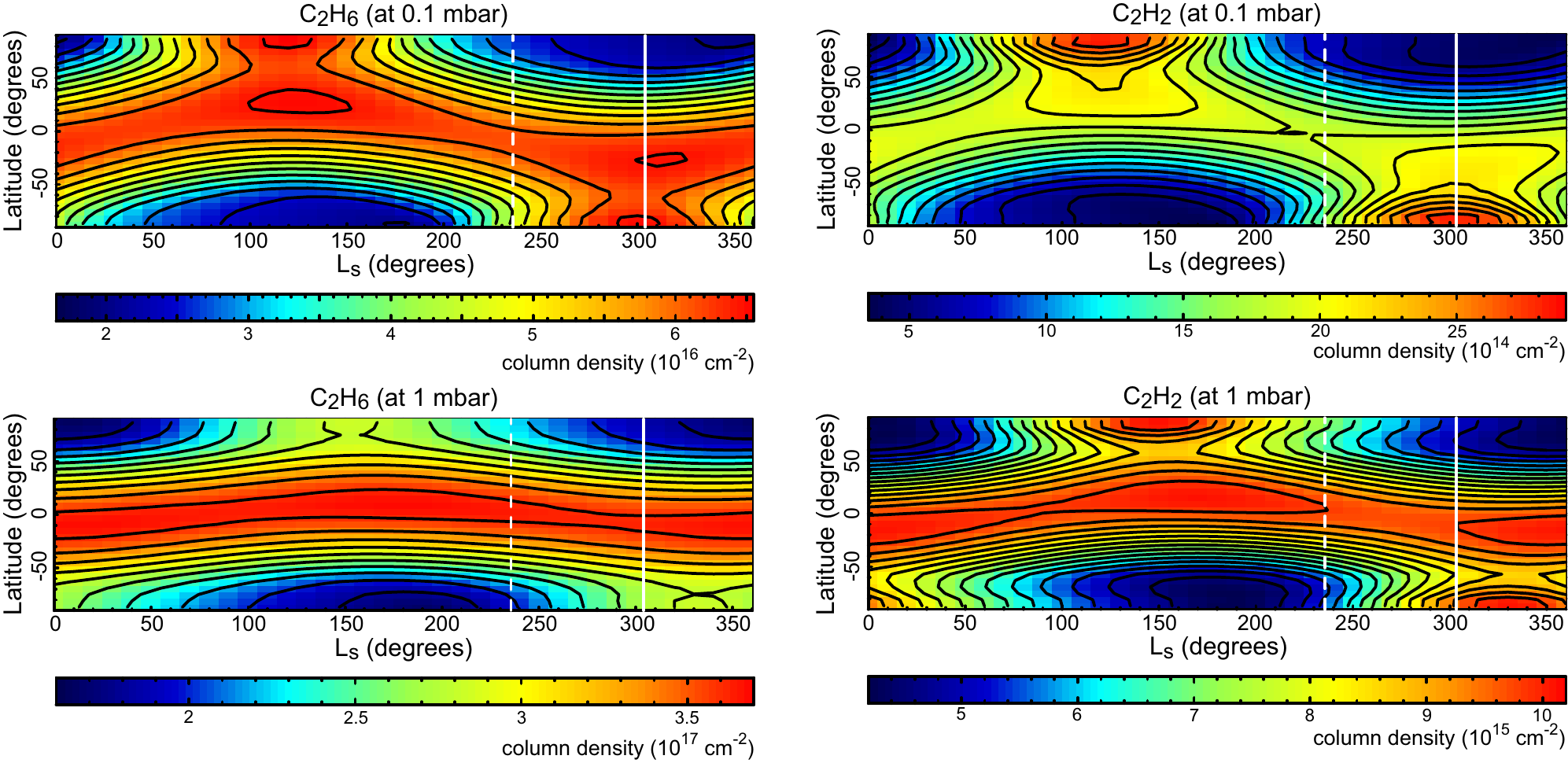}
\caption{Photochemical model predictions \cite{moses18} for the column abundance of C$_2$H$_6$ (Left) and C$_2$H$_2$ (Right) above 0.1 mbar (Top) and 
1 mbar (Bottom) on Neptune as a function of planetocentric latitude and season, where the season is represented by solar longitude $L_s$ ($L_s$ = 0$^\circ$ is 
northern vernal equinox, $L_s$ = 90$^\circ$ is northern summer solstice, etc.).  The white dashed line shows the season at the time of \textit{Voyager 2} 
encounter, and the white solid line shows the season at the current expected launch date (March 30, 2021) for the \textit{James Webb Space Telescope (JWST)}. 
Figure modified from Moses et al. \cite{moses18}.}
\label{fignepseas}
\end{figure}

Figure \ref{fignepseas} illustrates how the column abundance of two photochemically produced hydrocarbons, C$_2$H$_2$ and C$_2$H$_6$, are expected 
to vary with latitude and season on Neptune from the Moses et al. \cite{moses18} model.  At very high altitudes (not shown), chemical and transport 
time scales are short, and the atmosphere responds quickly to changes in the solar actinic flux.  The greater flux during the summer season leads 
to greater hydrocarbon production rates, and a generally 
higher abundance of photochemical products.  Conversely, hydrocarbon abundances drop off significantly during the long, dark winter at
high latitudes.  Seasonal variations in abundance are greatest at low pressures, but are reduced at depth as a result of diffusion and 
chemical time scales that increase with increasing pressure.  The greater time constants at higher pressures cause phase lags in the response 
to seasonal solar forcing, as can be seen in Fig.~\ref{fignepseas}.  Similarly, the meridional distribution of photochemical 
products becomes more symmetric about the equator at pressures greater than a few millibars  --- time constants that are 
longer than a Neptune season cause the annual average solar actinic flux to be more important in controlling abundances than the time-variable 
seasonal forcing at these pressures.  Seasonal differences in mixing ratios typically disappear at the few mbar level on Neptune or by $\sim$1 
mbar on Uranus, although these values depend on latitude and the species in question \cite{moses18}.

Surprisingly, seasonal variations in photochemical product abundances are expected to be weaker on Uranus than on Neptune \cite{moses18}, despite 
the greater obliquity of Uranus.  These weaker seasonal variations result from the weaker stratospheric vertical mixing on Uranus that confines methane 
to relatively deep pressure levels in the stratosphere.  Both the vertical transport time scales and chemical time scales are long in the 
region where CH$_4$ is photolyzed on Uranus, so the hydrocarbon photochemical products do not respond quickly to the variable solar forcing.  However, 
in the 0.1--1 mbar region that is probed by mid-infrared observations, some seasonal variations are expected, leading to north vs.~south hemispheric 
dichotomies in the photochemical product abundances during most seasons \cite{moses18}.  Species such as C$_2$H$_2$, C$_3$H$_4$, and C$_4$H$_2$ 
are predicted to exhibit a maximum at the poles in the summer-to-fall hemisphere due to both the annual-average solar insolation being greater at high latitudes 
than at low latitudes and to the phase lags in the response to the seasonally variable insolation.  In contrast, C$_2$H$_6$ is predicted to have a maximum at 
low latitudes because loss by photolysis effectively competes with production in the high-latitude summer.  

For Saturn, comparisons between the predicted and observed meridional distributions of temperatures and hydrocarbon abundances helped identify 
certain characteristics of stratospheric transport, such as regions of local 
upwelling and downwelling that can help define stratospheric circulation patterns
\cite{greathouse05,mosesgr05,howett07,fouchet08,hesman09,guerlet09,guerlet10,fletcher07,fletcher08hex,fletcher09midir,fletcher10,fletcher12beacon,fletcher15satpole,sinclair13,sinclair14,sylvestre15,hue15,moses15}.
The same could be true for Uranus and Neptune.  
Observations that define the meridional variations in photochemical product abundances have recently become available for Uranus \cite{roman20} 
and Neptune \cite{greathouse11,fletcher14nep}.  
The meridional distributions of C$_2$H$_2$ and C$_2$H$_6$ predicted by the Neptune photochemical model of Moses et al. \cite{moses18} 
appear to compare well to the limited available observations \cite{greathouse11,fletcher14nep}, suggesting that stratospheric 
transport has a more minor effect on species abundances at Neptune than it does at Saturn.  Further observations that map out species abundances as a 
function of altitude, latitude, and time are needed to confirm this claim.  

In contrast, Uranus exhibits strong changes in brightness temperature with latitude at 
13 $\mu$m \cite{roman20} that are not predicted by the seasonal model.  Under the assumption that the brightness-temperature variations result 
solely from changes in the mixing ratio of C$_2$H$_2$ rather than variations in temperature (i.e., recall the degeneracy between temperatures and abundances 
in analyses of emission data), the C$_2$H$_2$ mixing ratio at $\sim$0.2 mbar has a minimum at the equator in observations from 2009 ($L_s$ $\approx$ 7$^{\circ}$), 
with a maximum at mid-latitudes in both the northern and southern hemispheres \cite{roman20}.  That pattern does not change much in observations acquired in 
2018 ($L_s$ $\approx$ 46$^{\circ}$), where the equatorial minimum and northern mid-latitude local maximum are still seen (southern mid-latitudes  
are no longer in view) \cite{roman20}.  The inferred C$_2$H$_2$ abundances at mid- and high-latitudes are roughly consistent with model predictions, but the 
observed variations with latitude are far greater than predicted, and the equatorial minimum and local maximum at northern mid-latitudes are unexpected.

This mid-latitude peak in emission 
could potentially be produced by a local maximum in stratospheric temperatures, perhaps caused by a downwelling with consequent adiabatic heating.  However,
retrievals of upper-tropospheric temperatures suggest a tropospheric upwelling at these same mid-latitudes, so the stratospheric circulation would have to 
be in the opposite sense to that in the troposphere.  Roman et al. \cite{roman20} put forth an alternative explanation, suggesting that the relative 
enhancement at northern mid-latitudes could be caused by stronger stratospheric mixing at these latitudes (due to the tropospheric upwelling) that carries 
more methane to higher altitudes, allowing more C$_2$H$_2$ to be photochemically produced at these latitudes compared to the equatorial region, especially as 
the solar actinic flux is beginning to increase as summer approaches.  Preliminary modeling suggests that such enhancements would require considerable 
changes to the column abundances and vertical profiles of all hydrocarbons at these latitudes.  Regardless of whether either explanation 
is responsible, the discrepancy between current models and observations of Uranus is intriguing and might provide useful insight to coupled tropospheric-stratospheric 
dynamics and coupled stratospheric dynamics and chemistry.  To make further advances in this topic, an additional way to break the strong degeneracies 
between temperatures and abundances in mid-infrared emission observations would be valuable (e.g., spatially resolved measurements of the S(1) quadrupole 
line of H$_2$ --- as will be possible with \textit{JWST} \cite{norwood16} --- to help constrain stratospheric temperatures, \textit{in situ} temperature 
measurements from probe(s), better derivations of stratospheric CH$_4$ vertical and meridional distributions at visible and near-infrared wavelengths that 
are not so degenerate with temperature), as would additional observations that target the stratospheric distributions of hydrocarbons.  

\subsection{Effects of external material}

Oxygen-bearing species such as H$_2$O, CO$_2$, and CO are present in the stratospheres of Uranus and 
Neptune \cite{feuchtgruber99,lellouch05,hesman07,meadows08,fletcher10akari,teanby13,orton14chem,luszczcook13,cavalie14,teanby19}.  Any CO$_2$ and H$_2$O 
being carried up from the deep troposphere on Uranus and Neptune would be expected to condense long before reaching the stratosphere, and although 
CO does not condense and can have a deep tropospheric source, the fact that the mixing ratio of CO is greater in the stratosphere than the troposphere 
on both planets indicates a source from outside the planet.  Delivery of oxygen from the ablation of interplanetary dust particles is the right order of 
magnitude to explain the observed amount of H$_2$O, CO$_2$, and CO in the stratosphere of Uranus \cite{poppe16,moses17poppe}, but cometary impacts or local 
satellite/ring material could also contribute \cite{cavalie14,moses17poppe,lara19}.  For Neptune, the expected dust influx rates are far too small to 
explain the large observed amount of CO in Neptune's stratosphere, pointing to a very large cometary impact within the last $\sim$1000 years 
\cite{lellouch05,hesman07,luszczcook13,poppe16,moses17poppe,moreno17,dobrijevic20}.  The large CO amount, plus
its greater mixing ratio in the stratosphere than the troposphere, the large inferred stratospheric CO/H$_2$O ratio, and the observed presence of 
stratospheric hydrogen cyanide (HCN) \cite{marten93,rosenqvist92,guilloteau93,lellouch94,marten05} 
originally led Lellouch et al. \cite{lellouch05} to suggest that the CO was delivered to Neptune through a large cometary impact a few hundred years
ago.  Recent carbon monosulfide (CS) observations \cite{moreno17} strengthen this cometary-impact possibility.

Depending on the source of the external material, the oxygen species will be delivered to Uranus and Neptune at different pressure levels.  For gas 
coming in from a local satellite or ring source, the oxygen can flow in from the top of the atmosphere, affecting chemistry throughout the atmosphere.
For an interplanetary dust source \cite{poppe16}, ablation of icy grains releases oxygen-bearing species to the $\sim$10$^{-1}$ to $10^{-7}$ mbar 
region of Uranus and Neptune \cite{moses92abl,moses17poppe}, which affects chemistry both above and below the methane homopause.   Observations and 
models of the impacts of Comet Shoemaker-Levy 9 with Jupiter taught us a lot about how comet-derived material ends up in planetary stratospheres
\cite{zahnle96,lellouch96,moses96}.  
During a large cometary impact, vaporized cometary material plus some ambient gas from the terminal explosion deeper in the atmosphere will rush 
back up the entry column to form a plume of material that rises above the atmosphere before falling back upon the stratosphere \cite{boslough95,harrington04}.  
During this plume splashback phase, the material that re-enters the atmosphere is reshocked (resetting the molecular composition) and deposited within the 
middle stratosphere \cite{zahnle96,lellouch96,lellouch97,harrington04}.  The plume re-entry shock from the Shoemaker-Levy 9 impacts was characterized 
by relatively low shock pressures and high temperatures, with the maximum shock temperatures depending on the plume re-entry vertical velocity (and 
thus distance from the impact site) \cite{zahnle96}.  The fact that CO was favored over H$_2$O at the impact sites suggested that typical peak 
re-entry shock temperatures were above $\sim$1400 K during the Shoemaker-Levy 9 impacts \cite{zahnle96,lellouch96}.  In general, CO was found to be 
a factor of $\sim$10--100 more abundant that H$_2$O in the Jovian stratosphere after the impacts \cite{lellouch96,lellouch97,lellouch02}, and the CO, 
HCN, and CS that were formed during the plume splashback phase have persisted for years in the Jovian stratosphere 
\cite{moreno03,moreno06,griffith04,lellouch06,iino16jup}, as was predicted from photochemical models \cite{moses95sulf,moses95nit,moses96}.  
Observations suggest that these species were introduced at pressures less than $\sim$0.1 mbar after the impacts, and have been diffusing slowly 
downward since that time \cite{lellouch95,lellouch97,moreno03}.  External species delivered to atmospheres of Uranus and Neptune from large cometary impacts 
are expected to have similar behavior, although the amount of material introduced and its initial vertical distribution will depend on the size of 
comet, as well as its entry velocity.

\begin{figure}[!ht]
\centering\includegraphics[width=5.2in]{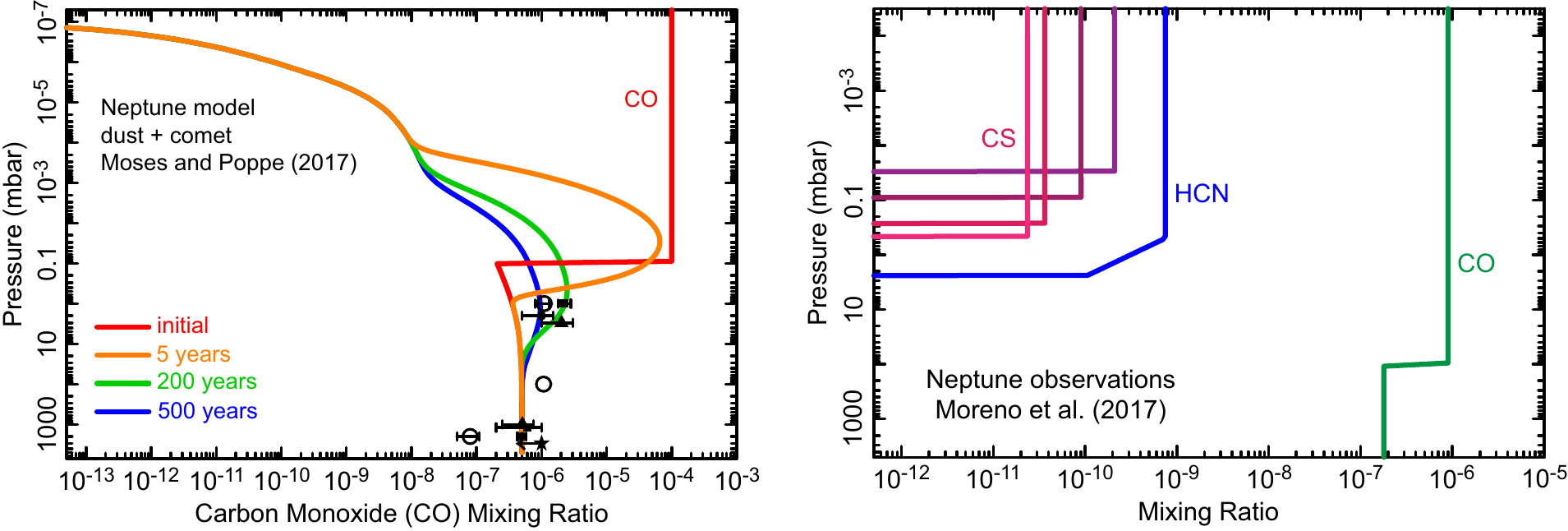}
\caption{(Left) Photochemical model predictions for the time evolution of CO introduced from a large cometary impact on Neptune, along with a smaller steady 
source from the ablation of icy interplanetary dust grains (Figure modified from Moses and Poppe \cite{moses17poppe}).  (Right) Mixing ratio profiles derived for 
CS, HCN, and CO from analyses of the millimeter and sub-millimeter observations of Moreno et al. \cite{moreno17} and Luszcz-Cook and de Pater 
\cite{luszczcook13}.  Note that although the column abundance of CS is well constrained from the observations, the mixing-ratio profile is not, so multiple models are 
shown that all provide a good fit to the data \cite{moreno17} (Figure modified from Moreno et al. \cite{moreno17}).}
\label{fignepcomet}
\end{figure}

Photochemical models can track the fate of the oxygen species introduced by these external sources 
\cite{moses92abl,moses05,orton14chem,cavalie14,moses17poppe,moses18,lara19,dobrijevic20} (see Fig.~\ref{fignepcomet}).  Potential sources from 
dust ablation or the local satellite/ring systems can be modeled as continuous sources, while cometary impacts require time-variable models.  In 
general, H$_2$O, CO, and CO$_2$ are relatively stable chemically in the cold stratospheres of the Ice Giants.  Chemical interactions with 
hydrocarbons do occur, but have only a minor effect on the abundances of the observable hydrocarbons.  See Moses et al. \cite{moses00b,moses05,moses17poppe}, 
Orton et al. \cite{orton14chem}, Lara et al. \cite{lara19}, and Dobrijevic et al. \cite{dobrijevic20} for a discussion of the photochemistry of oxygen 
species.  If delivered in sufficient amounts, externally supplied H$_2$O, CO$_2$, and HCN will condense in the stratospheres of Uranus and Neptune, 
contributing to a high-altitude haze at pressures ranging from 10$^{-2}$ mbar (e.g., H$_2$O on Uranus) to greater than 10 mbar (CO$_2$ on Neptune) 
(see Fig.~\ref{figchemurannep}).  

The vertical distributions 
of CO, HCN, and CS on Neptune differ significantly from each other \cite{moreno17}.  This interesting observation (see Fig.~\ref{fignepcomet}) 
could be the result of more than one large impact (with resulting vertical distributions that vary because of different ages or the size of the 
impactors), complicated and velocity-dependent plume splashback conditions, different rates 
of chemical loss for the different species, or a different external source altogether.  The HCN, for example, could potentially derive from chemistry 
resulting from the dissociation of quenched disequilibrium N$_2$ by galactic cosmic rays or by chemistry resulting from the inflow of 
nitrogen from Triton \cite{marten93,rosenqvist92,lellouch94,gautier95}.  The chemistry of nitrogen species in Neptune's atmosphere has been 
explored previously \cite{lellouch94}, 
albeit not in the context of a cometary impact, but the fate of sulfur species delivered by a comet on Neptune has not yet been studied.  The 
external material delivered to the Ice Giants could potentially affect ionospheric chemistry and structure, depending on the vertical distribution 
of the exogenic species \cite{lyons95,dobrijevic20}.

Note from Fig.~\ref{fignepcomet} that stratospheric CO from external sources is observed to extend down to at least 100 mbar pressure levels 
on Neptune \cite{luszczcook13,teanby19}, which suggests a very large impact occurred long enough ago ($\gta$ 1000 years) for the comet-deposited material 
to have descended all the way to the tropopause \cite{moreno17}; alternatively, perhaps the external CO results from a more continuous influx of 
small comets, as has been suggested for Jupiter \cite{bezard02}, or from local debris from the satellite/ring system.  These scenarios lead to the 
uncomfortable possibility that the large amount of CO observed in the troposphere of Neptune could have a contribution from external sources, which 
would complicate the use of CO as an indirect indicator of the deep oxygen abundance.  

\begin{figure}[!ht]
\centering\includegraphics[width=5.2in]{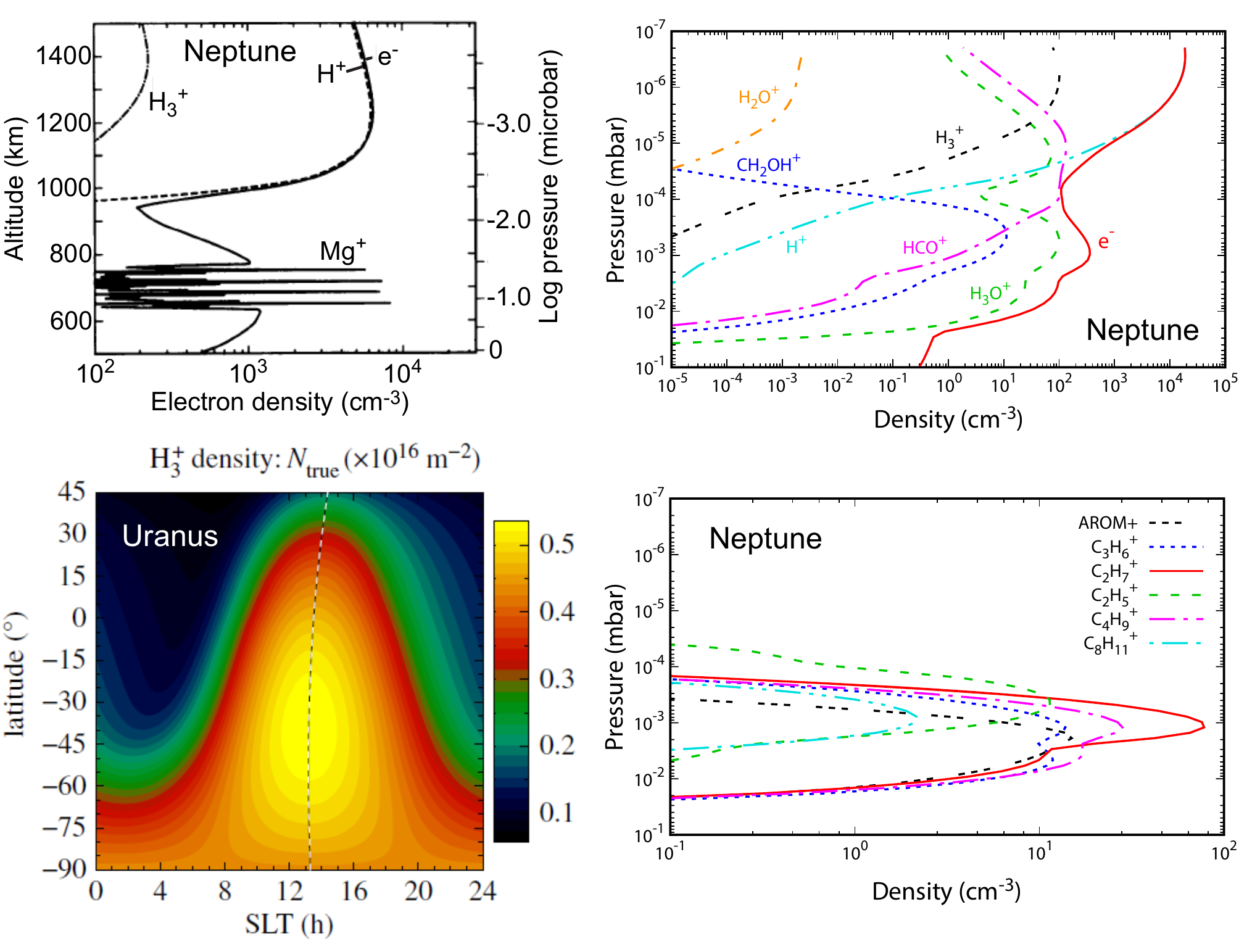}
\caption{Ion chemistry on Uranus and Neptune: 
(Top Left) Photochemical model for Neptune's ionosphere, in which Mg$^+$ ions form sharp layers in the lower ionosphere as a result of 
sinusoidal winds from a hypothetical atmospheric wave; figure modified from \cite{lyons95}.  (Bottom Left) Global photochemical model results 
for Uranus, illustrating the H$_3$$^+$ column 
density as a function of latitude and solar local time (SLT); figure from \cite{moore19}. (Top Right) Photochemical model results for Neptune, 
illustrating the vertical profiles of major hydrogen and oxygen ions; figure from \cite{dobrijevic20}.  (Bottom Right).  Photochemical model results 
for Neptune, showing the dominant hydrocarbon ions in the lower ionosphere; figure from \cite{dobrijevic20}.} 
\label{figion}
\end{figure}

\section{Chemistry of the thermosphere and ionosphere}

An ionosphere can form in the thermosphere and upper stratosphere of Uranus and Neptune from the interaction of the atmosphere with extreme ultraviolet 
radiation and X-rays.  Galactic and solar cosmic rays can also ionize species down to tropospheric levels \cite{capone77}.  Ionospheric models of Uranus and 
Neptune have been developed by several groups \cite{atreya75,atreya75hc,capone77,atreya83,chandler86,shinagawa89,lyons95,melin18,moore19,dobrijevic20}.  
Above the methane homopause at the top of the stratosphere, the ionosphere is dominated by H$^+$ and H$_3$$^+$, which form the main ionospheric peak; 
some amount of H$_3$O$^+$ and HCO$^+$ is expected to be present, as well, depending on the external oxygen source and vertical profile.  A 
secondary ionospheric peak forms in the lower thermosphere and upper stratosphere, populated by hydrocarbon ions (see Fig.~\ref{figion}).  Ablated refractory debris 
from micrometeoroids could deliver metal vapors \cite{moses92abl}, whose long-lived atomic ions could potentially replace hydrocarbon ions in this 
secondary peak \cite{lyons95}.  In fact, Lyons \cite{lyons95} suggests that sharp electron-density layering in the lower ionosphere of Neptune and the 
other giant planets, as seen from the \textit{Voyager} radio occultation experiment, could be caused by Mg$^+$ or other long-lived ions being 
compressed by horizontal winds with vertical shears (e.g., gravity waves) acting in the presence of the planet's magnetic field (see also 
\cite{mosesbass00,matcheva01}).  

Of all the ionic species predicted to be present on the Ice Giants, only H$_3$$^+$ has been detected, and only on Uranus
\cite{trafton93,trafton99,lam97,encrenaz03,feuchtgruber03,melin11uran,melin11nep,melin13,melin18}.  Observations of H$_3$$^+$ have provided 
a useful probe of thermospheric conditions on Uranus.  For example, thermospheric temperatures derived from H$_3$$^+$ observations have 
been observed to vary with time \cite{melin11uran,melin13}. The long-term trends in thermospheric temperature variations are interesting and 
appear to be real.   Moore et al. \cite{moore19} caution that solar-zenith angle effects resulting from the changing seasonal geometry 
of the planet, combined with temperature gradients and H$_3$$^+$ abundances that change with altitude, latitude, and local time, can 
influence derivations of thermospheric temperatures.  Such global considerations should be taken into account when interpreting H$_3$$^+$ 
observations.  For the case of Neptune, ionospheric models typically predict H$_3$$^+$
column densities greater than the observational upper limit \cite{melin18}.  The reasons for this discrepancy are not clear but may have to 
do with cooler thermospheric temperatures than were seen during the \textit{Voyager} encounters,  a larger than expected influx of external 
material interacting with the hydrogen ions and reducing their densities in favor of species like H$_3$O$^+$ or HCO$^+$, or a methane 
homopause level that is higher in the atmosphere than was observed during the \textit{Voyager} era (see Moore et al., this volume).

Dobrijevic et al. \cite{dobrijevic20} predict strong chemical coupling between ions and neutrals in the upper stratosphere of Neptune.  
They find that the production of neutral benzene and other aromatic hydrocarbons is enhanced several orders of magnitude by coupled 
ion-neutral chemistry, while C$_2$H$_6$ is unaffected, and the abundance of C$_2$H$_2$ is decreased by a factor of $\sim$1.5.  Neutral oxygen species such 
as CO$_2$ are also affected by ion chemistry \cite{dobrijevic20}.  These predictions remain to be tested observationally ---  ultraviolet 
occultations (e.g., \cite{koskinen16}) and mid-infrared limb spectra (e.g., \cite{guerlet15}) from an orbiting spacecraft could could 
provide useful tests of such models.

\section{Conclusion and summary of outstanding science questions}

Observations and models to date have revealed some crucial insights about atmospheric composition on Uranus and Neptune, but there is much 
we still do not know about atmospheric chemistry on our solar system's Ice Giants.  Several photochemical products have been detected, along
with a few equilibrium ``parent'' molecules such as H$_2$, CH$_4$, and H$_2$S, but we have little information about the vertical and horizontal
distributions of atmospheric constituents.  Information about temporal variability is equally limited.  
Many elements are tied up in condensates at depths, preventing direct determinations of the bulk elemental composition.  
Degeneracies in observational analyses --- such as between temperatures and constituent abundances in thermal emission observations (at wavelengths 
from the infrared to microwave) and between constituent abundances and aerosol extinction in reflected sunlight observations (at ultraviolet, 
visible, and near-infrared wavelengths) --- hamper the determinations of species mixing ratios on Uranus and Neptune.

When planning future Ice Giant missions, investigators should keep in mind the key outstanding science questions related to atmospheric chemistry on 
Uranus and Neptune, and the ways in which these questions can be addressed:

\begin{itemize}
\item \textbf{What is the elemental composition of the deep atmospheres of Uranus and Neptune, and what does that tell us about planetary 
formation processes?}  
Determining the elemental composition of the deep atmosphere can help constrain the composition and properties of the gas and 
solid material accreted by the planet, helping to 
distinguish between competing theories of planetary formation and evolution \cite{mousis18}.  Measurements of noble gases and their isotopes from \textit{in situ} 
probes would be extremely valuable in this regard, as would obtaining vertical abundance profiles of CH$_4$, NH$_3$, H$_2$S, and PH$_3$ down 
to tens of bars (and firm detections of NH$_3$ and PH$_3$ in the first place), and isotopic ratios of the main C, N, O carriers \cite{mousis18}.  
Measurements of the vertical profiles of disequilibrium quenched species such as CO, N$_2$, and potentially CO$_2$ and PH$_3$ would also help constrain 
the deep oxygen and nitrogen abundance --- a probe is unlikely to survive to deep enough pressures to measure the deep oxygen abundance on Uranus and 
Neptune.  Microwave or radio observations from an orbiter or from the Earth can remotely sample deeper tropospheric levels, albeit with coarse vertical 
resolution and potential degeneracies in contributions from various opacity sources and temperatures, to help constrain the deep abundance of sulfur, 
nitrogen, and oxygen, as well as put any probe measurements in a more global 3D context.  
\vskip6pt

\item \textbf{How does the tropospheric composition vary in three dimensions across Uranus and Neptune, and what does that tell us about 
convection, circulation, and dynamics; what are the implications for determining deep elemental abundances?}  
Condensible tropospheric gases on the giant planets are observed to experience significant and 
unexpected variations in mixing ratio with latitude and altitude \cite{bolton17sci,li17juno,li20,karkoschka09,karkoschka11,tollefson19}.  Determining 
how CH$_4$, H$_2$S, NH$_3$, and H$_2$O on Uranus and Neptune vary in a three-dimensional sense will help us better 
understand atmospheric circulation and convective processes in hydrogen-dominated atmospheres for which the condensate is heavier than the 
background gas (including on extrasolar planets).  
Determining the 3D distribution of disequilibrium quenched species such as CO, CO$_2$, N$_2$, and PH$_3$ would also shed light on convective 
processes in the deep troposphere where quenching occurs.
Both remote sensing and \textit{in situ} probe measurements can provide complementary information on tropospheric species distributions \cite{cavalie20}.  
Probes can provide direct measurements of the thermal structure of a specific region or regions, which would provide an invaluable means with which 
contemporaneous remote-sensing observations can be calibrated to help break degeneracies between temperatures and abundances.  Spatially resolved 
remote-sensing observations from ground-based or space-based facilities can help identify dynamically favorable entry regions for potential 
\textit{in situ} probes to maximize the likelihood of sampling the deep atmospheric abundances \cite{fletcher20circ}.
\vskip6pt

\item \textbf{Can disequilibrium chemical tracers such as CO, CO$_2$, H$_2$CO, C$_2$H$_6$, N$_2$, HCN, and PH$_3$ provide robust indirect 
indicators of the deep abundance of O, N, P, and other elements?}  Theoretical models suggest that these and other quenched disequilibrium 
species will have their mixing ratios frozen in when the rate of atmospheric transport exceeds that of key kinetic reactions in the deep troposphere; 
consequently, the observed upper tropospheric abundance can provide an indirect measure of the deep elemental abundances on Uranus and Neptune 
\cite{prinn77,fegley85uran,fegley91,luszczcook13,cavalie17,venot20fixed,cavalie20}.  However, uncertainties in these models need to be reduced 
before we can make meaningful constraints on deep elemental abundances through this process (see section 3a).  Moreover, the vertical profiles 
and latitude distribution of these species would need to 
be obtained and combined with models to accurately determine the effect of external material and potentially latitude-dependent deep convection.  
Both \textit{in situ} probes and remote-sensing observations would be useful in 
this regard, as would better constraints on vertical temperature profiles across the planet.  If a probe is used to determine disequilibrium 
constituent abundances, having some means to distinguish between CO and N$_2$ --- both with mass 28 amu, and both expected to be present in 
similar abundance --- would be valuable, as could be obtained with a high-resolution mass spectrometer or specific targeted instrumentation.
\vskip6pt

\item \textbf{How does the stratospheric composition vary in three dimensions across Uranus and Neptune, and what does that distribution tell 
us about atmospheric circulation, moist convection, atmospheric structure, chemical processes, response to seasonal forcing, and sources 
of external material?} The 3D distribution of stratospheric species depends on the source and distribution of the parent molecules, 
photochemical kinetics, seasonal variations in solar insolation, atmospheric transport, and aerosol microphysical processes.  Identifying 
how constituents are distributed across the planet would help quantify the relative importance of the different processes.  Spatially 
resolved mid-infrared spectroscopy from large ground-based telescope facilities can provide a means for mapping temperatures and 
photochemical-product abundances; however, such instrumentation on an Ice-Giant orbiter would 
provide better spatial resolution and coverage, particularly of winter hemispheres inaccessible from Earth.
Temperature measurements from \textit{in situ} probes would provide a valuable benchmark to calibrate such remote-sensing observations.  
Ultraviolet and near-infrared stellar and solar occultations could help identify new species and define how the methane homopause level varies across 
the planet, which in turn is important for understanding photochemical processes and coupled atmospheric dynamics and chemistry.  Ground-based 
sub-millimeter observations can be used to monitor the distribution of CO, HCN, and CS to help identify the external sources and measure 
stratospheric winds. Sub-millimeter instruments on an Ice Giant orbiter could additionally provide measurements of H$_2$O and perhaps 
H$_2$CO, CH$_3$OH, and CH$_3$C$_2$H.  Ultraviolet, visible, and 
near-infrared observations at a variety of phase angles can help constrain the aerosol structure and scattering properties, which in turn can 
help characterize the distribution and composition of potential condensible gases.  
Mapping of stratospheric methane via near- and mid-infrared observations can further our 
understanding of how methane is transported past the tropopause cold trap into the stratospheres of Uranus and Neptune.  Identifying 
how these processes work on our Ice Giants will help predict and understand exoplanet observations.
\vskip6pt

\item \textbf{To what extent is cometary or other external debris affecting the troposphere, stratosphere, thermosphere, and ionosphere of 
Uranus and Neptune, and what are the implications for impact rates in the outer solar system, atmospheric structure and chemistry,
and our ability to infer deep atmospheric abundances?}  External material is clearly supplying the upper atmospheres of our Ice Giants with 
oxygen, nitrogen, and sulfur-bearing species and potentially other elements that would not normally be present in the upper
atmosphere.  This external material can strongly affect ionospheric chemistry and structure, stratospheric chemistry and aerosol formation, and 
radiative properties of the atmosphere.  It is currently unclear whether this external material is contributing significantly to 
the CO abundance in the upper troposphere, which we need to resolve before we can use this potential quenched disequilibrium species as a 
tracer for the deep oxygen abundance on Uranus and Neptune.  Obtaining accurate vertical profiles of CO in the lower stratosphere and upper 
troposphere from either \textit{in situ} probes or remote-sensing observations that resolve line shapes will be critical for separating out 
the contributions from the internal and external sources of the CO.  For similar reasons, as well as to identify the source of the external 
material, obtaining vertical profiles and the spatial distribution of H$_2$O, CO$_2$, HCN, and CS would be valuable.  In the near term, 
observations from ALMA and \textit{JWST} can help provide some of this information.
\vskip6pt

\item \textbf{What is the composition and structure of the ionosphere of Uranus and Neptune, and what does that tell us about thermospheric 
processes, magnetospheric processes, and sources of external material?}  We currently have very limited information about the ionospheric 
structure and composition on Uranus and Neptune.  External material from the local satellite/ring systems, interplanetary dust particles, 
or cometary impacts can potentially affect the overall electron-density profile, chemical speciation, and layering within the ionosphere.  The 
extent to which the Ice Giants' unusual offset, tilted, complex magnetic field influences ionospheric properties and promotes 
magnetosphere-atmosphere interactions is not well understood, nor is how the ionosphere varies spatially and temporally.  Radio occultations 
from an Ice Giant orbiter would help map out ionospheric structure across the Ice Giants to help constrain ionospheric processes and the influence of 
external material.  Near-infrared emission observations from an Ice Giant orbiter or Earth-based observations can monitor H$_3$$^+$ to track spatial and 
temporal variations in the ionosphere and help search for additional ionic species.  Ultraviolet occultations could help map species such as 
benzene that are produced from ion-neutral chemistry and contribute to stratospheric haze deeper in the atmosphere.
\vskip6pt

\item \textbf{Why is the observable atmospheric composition of Uranus and Neptune so different, are these differences variable or static, 
and what are the implications for interior and deep atmospheric processes and atmospheric evolution?
How is the atmospheric composition of Uranus and Neptune affected by the planet's internal heat flow, relative abundance of 
different elements, axial tilt, and orbital distance, and what are the implications for extrasolar planets?}
Despite a similar mass, radius, and bulk composition, the observed atmospheric composition of Uranus and Neptune is very different.  The two 
planets may have started out similarly but evolved differently, perhaps as a result of cataclysmic giant impacts 
\cite{stevenson86,podolak12,kegerreis18,reinhart20}.  Differences in internal heat flux between the two planets appear to lead to differences 
in the way the energy is transported through the atmosphere (e.g., \cite{vazan20}), which in turn leads to 
differences in atmospheric convection, tropospheric storm generation, and stratospheric mixing.  Differences in planetary obliquity may also lead to 
differences in atmospheric chemistry and dynamics.  
To fully understand how different physical and chemical processes on the two planets have shaped their different atmospheric composition, and how 
that composition might be time-variable, 
we need the \textit{in situ} probe and Ice Giant orbiter remote-sensing observations described above, along with monitoring of the composition 
over seasonal time scales from astronomical facilities on the ground and in space, and related accompanying laboratory and theoretical investigations 
into atmospheric properties and processes.  Knowledge gained 
from studies of our Ice Giants should be considered from the more general context of how similar atmospheric chemical and physical processes 
apply to the diverse population of Neptune-class and sub-Neptune exoplanets within our galaxy. 
\vskip6pt
\end{itemize}

Our mysterious Ice-Giant planets formed differently from both the Gas Giants and terrestrial planets; Uranus and Neptune are our only nearby representatives 
of a class of planet that is now known to be common in our galaxy.  The atmospheric composition on Uranus and Neptune holds valuable clues to 
both solar-system formation processes and to physical and chemical processes that operate on intermediate-sized volatile-rich planets.  To fully 
interpret these clues, we need to better understand the chemistry of these worlds and the complex coupling of chemistry, dynamics, and 
radiation that shape the observable atmospheric composition.  Many of the key outstanding atmospheric chemistry questions discussed above 
cannot be answered through remote observations from the Earth, due to the vast distances involved and the restriction to the Earth-facing hemisphere 
--- a dedicated exploration mission is required.  The scientific motivation for a future Ice Giant mission has never been stronger, and these distant 
worlds are beckoning.  

\enlargethispage{20pt}

\dataccess{This article has no additional new data. The data presented in the figures can be obtained from the original journal and/or the authors 
of that originating article.}

\aucontribute{J.M.~led the project and wrote the paper.  T.C., L.F., and M.R.~provided comments, additions, and corrections to the paper.}

\competing{We declare no competing interests.}

\funding{J.M.~acknowledges support from the NASA Solar System Workings grant 80NSSC19K0536.  T.C.~acknowledges funding from CNES and from the 
Programme National de Plan{\'e}tologie (PNP) of CNRS/INSU. L.F.~was supported by a Royal Society Research Fellowship at the University of Leicester.  
L.F.~and M.R.~acknowledge support from a European Research Council Consolidator Grant (under the European Union's Horizon 2020 research and innovation 
programme, grant agreement No.~723890).}

\ack{We thank Emmanuel Lellouch and an anonymous reviewer for comments and suggestions that improved the manuscript.}




\vskip2pc

%
%
%
%
%
%
%
%

\bibliographystyle{RS}
\bibliography{references}

\end{document}